\documentclass[%
preprint,
showpacs,
nofootinbib,
 amsmath,amssymb,
 aps,
prd,
]{revtex4-1}

\usepackage{graphicx}
\usepackage{dcolumn}
\usepackage{bm}

\begin{document}


\title{Coupled tachyonic dark energy: a dynamical analysis}

\author{Ricardo C. G. Landim}
\email{rlandim@if.usp.br}
 \affiliation{%
 Instituto de F\'isica, Universidade de S\~ao Paulo\\
 Caixa Postal 66318,  05314-970 S\~ao Paulo, S\~ao Paulo, Brazil
}%



\date{\today}

\begin{abstract}
In this paper we present a dynamical analysis for a coupled tachyonic dark energy with dark matter. The tachyonic field $\phi$ is considered in the presence of barothropic fluids (matter and radiation)  and the autonomous system due to the evolution equations is studied. The three cosmological eras (radiation, matter and dark energy) are described through the critical points, for a generic potential $V(\phi)$.
\end{abstract}



\maketitle

\section{Introduction}

The role of the tachyon in theoretical physics has a long story. For the most part of it, tachyon has been considered an illness whose  treatment and cure have been hard to be achieved. In the field theory, tachyon is associated with a negative mass-squared particle, which means that the potential is expanded around its maximum point. The natural question that arises is  whether tachyon potential has a good minimum elsewhere. The mexican hat potential in the spontaneous symmetry breaking is an example of this perspective. In the bosonic string, tachyon appears in its ground state, whereas in supersymmetric string theory its appearence can be avoided. However, even in superstrings, tachyon is present in non-BPS Dp-branes.  However, the potential of such Dp-branes does have a minimum \cite{Sen:1998sm, Sen:1999xm} and at this minimum  the tachyon field behaves like a pressureless gas \cite{Sen:2002in}. 

As soon as tachyon condensation in string theory had been proposed, its role in cosmology was studied, and plenty of works have thenceforth been done. Concerning the observational evidence for the accelerated expansion of the universe \cite{reiss1998,perlmutter1999}, tachyon can be a way to explain it, as a dynamical dark energy candidate (see \cite{copeland2006dynamics} for review).  Still regarding a dynamical dark energy, there exists the possibility of interaction between dark energy and dark matter \cite{Wetterich:1994bg,Amendola:1999er}, since their densities are comparable. This approach was applied to tachyons as well, in \cite{Gumjudpai:2005ry,micheletti2009}.

When a scalar field is in the presence of a barothropic fluid (with equation of state $w_m=p_m/\rho_m$)  the relevant evolution equations can be converted   into an autonomous system. Such approach was done for uncoupled dark energy (quintessence, tachyon field, phantom field and dilatonic ghost condensate, for instance \cite{copeland2006dynamics}) and coupled dark energy \cite{Gumjudpai:2005ry,TsujikawaGeneral,AmendolaChallenges,ChenPhantom}.   Since dynamical systems theory is a good tool to analyze asymptotic states of cosmological models, its usage  to study coupled tachyonic dark energy is in order. General analyses were pointed out in  \cite{TsujikawaGeneral,AmendolaChallenges} and the case with the potential $V(\phi)\propto \phi^{-2}$ was done in  \cite{Gumjudpai:2005ry}. In this paper, we enlarge the previous possibilities considering a generic potential and we assume a new form for the interaction between the two components of the dark sector, which in turn leads to new fixed points. Then, we solve the dynamical system and we analyze the stability of the fixed points, to check what kind of dominated universe can come up. Finally, we point out some specific potentials to illustrate the results.

The rest of the paper is organized in the following manner. In Section \ref{tachdyn} we present the dynamics of the tachyon field, in Section \ref{autosystem} we analyze the system in the presence of  barothopic fluids, through the autonomous system. Its fixed points are found and their stability analyzed. Section \ref{conclu} is reserved for the summary. We use Planck units ($\hbar=c=1 =M_{pl}=1$) throughout the text.

\section{Tachyon dynamics}\label{tachdyn}

\subsection{Basics}
 
 Tachyon field $\phi$ is described by the Born-Infeld Lagrangian
 
\begin{equation}\label{LBI}
 \mathcal{L}_{BI}=-\sqrt{-g}V(\phi)\sqrt{1-\partial^\mu\phi\partial_\mu\phi},
\end{equation} 

\noindent where $V(\phi)$ is the tachyon potential. The equation of motion for the tachyon field is

\begin{equation}\label{}
 -\nabla^\mu\partial_\mu\phi+\frac{\nabla_\mu\partial_\nu\phi}{1+\partial^\mu\phi\partial_\mu\phi}\partial_\mu\phi\partial_\nu\phi+\frac{V'(\phi)}{V(\phi)}=0,
\end{equation} 

\noindent For a homogeneous field, in an expanding universe with Friedmann-Robertson-Walker metric, the equation of motion becomes

\begin{equation}\label{eqmotion}
 \frac{\ddot{\phi}}{1-\dot{\phi}^2}+3H\dot{\phi}+\frac{V'(\phi)}{V(\phi)}=0.
\end{equation}

\noindent where the prime denotes time derivative with respect to $\phi$. The energy-momentum tensor for the tachyon field is

 \begin{equation}\label{Tmunu}
 T_{\mu\nu}=\frac{V(\phi)\partial_\mu\phi\partial_\nu\phi}{\sqrt{1+\partial^\alpha\phi\partial_\alpha\phi}}-g_{\mu\nu}V(\phi)\sqrt{1+\partial^\alpha\phi\partial_\alpha\phi},
\end{equation}

\noindent which leads to the energy density and pressure for the fluid given by

 \begin{equation}\label{energydens}
 \rho_\phi=\frac{V(\phi)}{\sqrt{1-\dot{\phi}^2}},
\end{equation}

 \begin{equation}\label{pressure}
 p_\phi=-V(\phi)\sqrt{1-\dot{\phi}^2}.
\end{equation}

With these quantities the equation of state for the dark energy becomes

\begin{equation}\label{eqstate}
 w_\phi=\frac{p_\phi}{\rho_\phi}=\dot{\phi}^2-1,
\end{equation}

\noindent thus, the tachyon behavior is between the cosmological constant one ($w_\phi=-1$) and matter one ($w_\phi=0$). Making use of Eqs.  (\ref{energydens}) and (\ref{pressure}) we get the second Friedmann equation 

\begin{equation}\label{}
\frac{\dot{a}}{a}=\frac{V(\phi)}{3\sqrt{1-\dot{\phi}^2}}\left(1-\frac{3}{2}\dot{\phi}^2\right).
\end{equation}

\noindent From the equation above we see that an accelerated expansion occurs for $\dot{\phi}^2<2/3$.

\subsection{Interacting dark energy}\label{de}

We now consider that dark energy and dark matter are coupled, but the total energy-momentum is still conserved. The continuity equations for both components are

\begin{equation}\label{contide}
\dot{\rho_\phi}+3H(\rho_\phi+p_\phi)=-\mathcal{Q},
\end{equation}

\begin{equation}\label{contimatter}
\dot{\rho_m}+3H\rho_m=\mathcal{Q},
\end{equation}

\noindent respectively, where $\mathcal{Q}$ is the coupling. In principle, the coupling  can depend on several variables $\mathcal{Q}=\mathcal{Q}(\rho_m,\rho_\phi, \dot{\phi},H,t,\dots)$, so we assume that $\mathcal{Q}=Q \rho_m\rho_\phi\dot{\phi}/H$, where $Q$ is a constant. With this form, the time dependence of the coupling is implicit in the Hubble parameter $H$. Thus, using (\ref{contide}), we get an equation of motion for the tachyon similar to (\ref{eqmotion}), but with an extra term due to the interaction 

\begin{equation}\label{eqmotioncoup}
 \frac{\ddot{\phi}}{1-\dot{\phi}^2}+3H\dot{\phi}+\frac{V'(\phi)}{V(\phi)}=-\frac{Q\rho_m}{H}.
\end{equation}

We are now going to analyze the  coupled tachyonic dark energy in the presence of radiation and, of course, matter.

\section{Autonomous system for the model}\label{autosystem}

 The Friedmann equations for the tachyon field in the presence of barothropic fluids are 
 
 \begin{equation}\label{eq:1stFEmatter}
  H^2=\frac{1}{3}\left(\frac{V(\phi)}{\sqrt{1-\dot{\phi}^2}}+ \rho_m+\rho_r\right),
\end{equation}

\begin{equation}\label{eq:2ndFEmatter}
  \dot{H}=-\frac{1}{2}\left(\frac{V(\phi)\dot{\phi}^2}{\sqrt{1-\dot{\phi}^2}} +\rho_m+\frac{4}{3}\rho_r\right),
\end{equation}

\noindent where the index $r$ stands for radiation. The  continuity equations are Eq. (\ref{eqmotioncoup}), for dark energy, Eq. (\ref{contimatter}) for matter and 

\begin{equation}\label{contirad}
\dot{\rho_r}+4H\rho_r=0,
\end{equation}

\noindent for radiation, with $H$ given now by Eq. (\ref{eq:1stFEmatter}). To deal with the dynamics of the system, it is convenient to  define the dimensionless variables \cite{Aguirregabiria:2004xd,Copeland:2004hq}

\begin{equation}\label{eq:dimensionlessXY}
\begin{aligned}
 x\equiv  \dot{\phi}, &\qquad y\equiv \frac{\sqrt{V(\phi)}}{\sqrt{3}H}, \qquad z\equiv\frac{\sqrt{\rho_r}}{\sqrt{3}H}\\ 
 &\lambda\equiv -\frac{V'}{V^{3/2}}, \qquad \Gamma\equiv \frac{VV''}{V'^2},\\
\end{aligned}
\end{equation}

\noindent which are going to characterize a system of differential equations in the form $\dot{x}=f(x,y,\dots,t)$, $\dot{y}=g(x,y,\dots,t)$, $\dots$, so that $f$, $g$, $\dots$ do not depend explicitly on time. For this system (called \textit{autonomous}), a point ($x_c$, $y_c$, $\dots$) is called \textit{fixed} or \textit{critical point}  if $(f, g,\dots)|_{x_c,y_c,\dots}=0$ and it is an \textit{attractor} when $(x(t),y(t),\dots)\rightarrow (x_c,y_c,\dots)$ for $t\rightarrow \infty$. 

The dark energy density parameter is written in terms of these new variables as

\begin{equation}\label{eq:densityparameterXY}
 \Omega_\phi \equiv \frac{\rho_\phi}{3H^2} = \frac{y^2}{\sqrt{1-x^2}},
 \end{equation}

\noindent so that Eq. (\ref{eq:1stFEmatter}) can be written as 

\begin{equation}\label{eq:SomaOmegas}
\Omega_\phi+\Omega_m+\Omega_r=1,
\end{equation}

\noindent where the matter and radiation density parameter are defined by $\Omega_i=\rho_i/(3H^2)$, with $i=m,r$. From Eqs. (\ref{eq:densityparameterXY}) and (\ref{eq:SomaOmegas}) we have that $x$ and $y$ are restricted in the phase plane by the relation

\begin{equation}\label{restriction}
0\leq x^2+y^4\leq 1,
 \end{equation}
 
\noindent due to $0\leq \Omega_\phi\leq 1$. In terms of these new variables the equation of state $w_\phi$  is

\begin{equation}\label{eq:equationStateXY}
 w_\phi =x^2-1,
\end{equation}

\noindent and the total effective equation of state is

\begin{equation}\label{eq:weff}
 w_{eff} = \frac{p_\phi+p_r}{\rho_\phi+\rho_m+\rho_r}=-y^2\sqrt{1-x^2}+\frac{z^2}{3},
\end{equation}

\noindent An accelerated expansion occurs for  $w_{eff} < -1/3$.

 The equations of motion for the variables $x$, $y$, $z$  and $\lambda$ are obtained taking the derivatives  with respect to $N\equiv \log a(t)$, where $a(t)$ is the scale factor and we set the present scale factor $a_0$ to be one,

\begin{subequations}\label{dynsystem}\begin{align}\label{eq:dx/dn}
\frac{dx}{dN}=(x^2-1)\left[3x-\sqrt{3}y\lambda+3Q\left(1-z^2-\frac{y^2}{\sqrt{1-x^2}}\right)\right],\end{align}

\begin{align}\label{eq:dy/dn}
\frac{dy}{dN}=\frac{y}{2}\left[-\sqrt{3}xy\lambda+ \frac{3y^2}{\sqrt{1-x^2}}(x^2-1)+3+z^2\right],
\end{align}

\begin{align}\label{eq:dz/dn}
\frac{dz}{dN}=-2z+\frac{z}{2}\left[\frac{3y^2}{\sqrt{1-x^2}}(x^2-1)+3+z^2\right],
\end{align}

\begin{eqnarray}\label{eq:dlambda/dn}
\frac{d\lambda}{dN}=-\sqrt{3}\lambda xy\left(\Gamma-\frac{3}{2}\right).
\end{eqnarray}
\end{subequations}

\subsection{Critical points}

 The fixed points of the system are obtained by setting $dx/dN=0$, $dy/dN=0$, $dz/dN$ and $d\lambda/dN=0$ in (\ref{dynsystem}). When $\Gamma=3/2$, $\lambda$ is constant the potential has the form found in  \cite{Aguirregabiria:2004xd,Copeland:2004hq} ($V(\phi)\propto \phi^{-2}$), known in the literature for both coupled \cite{Gumjudpai:2005ry,micheletti2009} and uncoupled  \cite{Padmanabhan:2002cp,Bagla:2002yn} dark energy. The fixed points are shown in Table \ref{criticalpoints}. Notice that $y$ cannot be negative and recall that $\Omega_r=z^2$. 
 
 \begin{table}[ph]
\caption{Critical points ($x$, $y$, $z$ and $\lambda$) of the Eq. (\ref{dynsystem}). The table shows the correspondent equation of state for the dark energy (\ref{eq:equationStateXY}), the effective equation of state (\ref{eq:weff}) and the density parameter for dark energy (\ref{eq:densityparameterXY}), for raditation ($\Omega_r=z^2$) and for matter from Eq. (\ref{eq:SomaOmegas}).}
{\begin{tabular}{@{}cccccccccc@{}} \toprule
Point & $x$ & $y$ & $z$&$\lambda$& $w_\phi$ & $\Omega_\phi$& $\Omega_r$&$\Omega_m$ & $w_{eff}$ \\
 \colrule
(a)         & $\pm 1$ & 0& 0&any value& 0& 0&0&1&0  \\
  (b) & $\pm 1$ & 0 &$\pm 1$ &any value&0 &0 &1&0&$\frac{1}{3} $\\
   (c) &0 & 0& $\pm 1$&any value& $-1$& 0& 1& 0& $\frac{1}{3}$\\
  (d) & 0 & 1 & 0& 0& $-1$ &1 &0 &0 & $-1$\\
  (e) & $\frac{\lambda y_c}{\sqrt{3}}$ & $y_c$ & 0& constant or 0 & $\frac{\lambda^2 y_c^2}{3}-1$&1  & 0& 0& $w_\phi$\\ 
  (f) & $-Q$ & 0 &0 &any value&$Q^2-1$ &0 &0 &1 & 0\\
  (g) & $x_f$ & $y_f$ &0 & constant and $\lambda>0$ &$x_f^2-1$ &$\frac{w_{eff}}{w_\phi}$ & 0&$1-\frac{w_{eff}}{w_\phi}$ &$\frac{x_fy_f\lambda}{\sqrt{3}}-1$ \\ 
  &  &  & & for $Q>0$ or  $\lambda<0$ for $Q<0$& & & & & \\
  \botrule
\end{tabular} \label{criticalpoints}}
\end{table}

The fixed points $y_c$, $x_f$ and $y_f$ are shown below

\begin{equation}\label{yc}
y_c=\sqrt{\frac{\sqrt{\lambda^4+36}-\lambda^2}{6}},
\end{equation}

\begin{equation}\label{xf}
x_f=-\frac{Q}{2}\pm \frac{\sqrt{Q^2+4}}{2},
\end{equation}

\begin{equation}\label{yf}
y_f=\frac{-\lambda x_f+ \sqrt{\lambda^2x_f^2+12\sqrt{1-x_f^2}}}{\sqrt{12(1-x_f^2)}}.
\end{equation}

The point (a) corresponds to a matter-dominated solution, since $\Omega_m=1$ and $w_{eff}=0$.  Points (b) and (c) are radiation-dominated solutions, with $\Omega_r=1$ and $w_{eff}=1/3$. The difference between them is that (b) has $w_\phi=0$, while (c) has $w_\phi=-1$, which means that there is another difference due to their stability, as we will see in the next section. Point (d) is a dark-energy-dominated solution with $\Omega_\phi=1$ and $w_{eff}=w_\phi=-1$. Point (e) is also a dark-energy-dominated solution ($\Omega_\phi=1$) but now the equation of state depends on $\lambda$, which in turn can be either constant or zero. The case with constant $\lambda$ is shown in the Table \ref{criticalpoints} and an accelerated expansion occurs for $\lambda^2<2/\sqrt{3}$.  For $\lambda=0$ we have $y_c=1$, $x_c=0$ and $w_\phi=-1$, thus we recover the point (d). All the points shown so far, but (d),  were also found in  \cite{Gumjudpai:2005ry,Aguirregabiria:2004xd,Copeland:2004hq} and they  do not have any dependence on $Q$, although the stability of the point (e) does, for the interaction showed here. Point (f) is also a matter-dominated solution ($\Omega_m=1$ and $w_{eff}=0$), but the equation of state for dark energy is $w_\phi=Q^2-1$, leading to an universe with accelerated expansion for $Q^2<2/3$. For this point the coupling is restrict to values $0\leq Q^2\leq 1$.

\begin{figure}[pb]
\centering
\includegraphics[width=9cm]{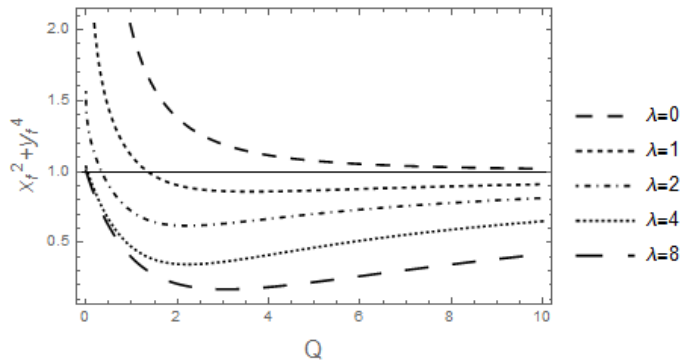}
\vspace*{8pt}
\caption{Restriction $x_f^2+y_f^4$ as a function of the coupling $Q$, for $\lambda=0,1,2,4$ and $8$. The horizontal  line shows the upper bound for the restriction. Thus, the fixed point does not exist for some values of $\lambda$ and $Q$. \label{f1}}
\end{figure}

\begin{figure}[pb]
\centering
\includegraphics[width=9cm]{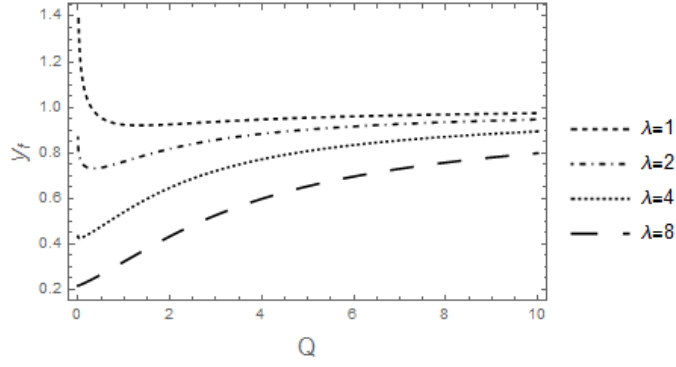}
\vspace*{8pt}
\caption{Fixed point $y_f$ as a function of the coupling $Q$, for $\lambda=1,2,4$ and $8$. \label{f2}}
\end{figure}

The last fixed point (g) requires more attention. This solution is valid for $x_f\neq 0$\footnote{The case for $x_f=0$ is the fixed point (d).} and for $Q\neq 0$, and its behavior depends on $Q$. In order to have $x^2_f\leq1$, we must have $Q>0$ for the case with plus sign in $x_f$ (\ref{xf}), while we have $Q<0$ for the minus sign case. When $Q\rightarrow |\infty|$\footnote{$Q\rightarrow +\infty$ for the plus sign in $x_f$ and $Q\rightarrow -\infty$ for the minus sign in $x_f$.}, $x_f\rightarrow 0$ and $y_f\rightarrow 1$, as it should be, from the restriction (\ref{restriction}). Furthermore, the fixed point exists for some values of $\lambda$ and $Q$, due to Eq. (\ref{restriction}). In Figure \ref{f1}, we show the restriciton  $x_f^2+y_f^4$ as a function of $Q$, for some values of $\lambda$. As we see, for $\lambda=0$ we have $y_t=1$ only when $x_f\rightarrow 0$, but this value is never reached since the fixed point (g) is valid for $x_f\neq 0$. Thus, for $\lambda\leq 0$ the restriction is not satisfied. From the figure we also notice that there is a lower bound for $Q$, for which the point (g) start existing. In Figure \ref{f2} we plot $y_f$ as a function of $Q$, for the all the values of $\lambda$, but zero, shown in Figure \ref{f1}.\footnote{We used the plus sign in $x_f$ (\ref{xf}) for Figures \ref{f1} and \ref{f2}, but the minus sign shows similar behavior, with $Q<0$ due to Eq. (\ref{restriction}),  satisfied for  $\lambda<0$. The graphics for this case are reflections in the ordinate axis. }

\begin{figure}[pb]
\centering
\includegraphics[width=9cm]{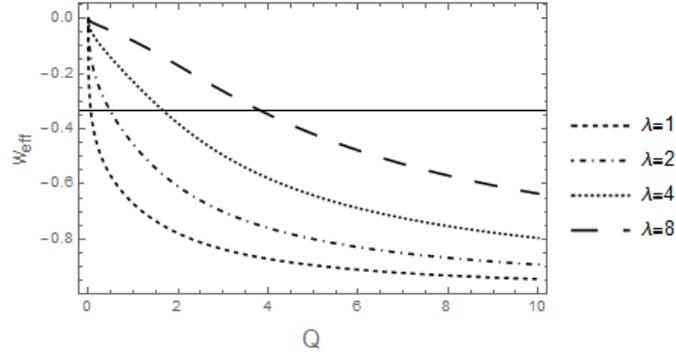}
\vspace*{8pt}
\caption{Effective equation of state for the fixed point (g) as function of $Q$ for  $\lambda=1,2,4$ and $8$. The horizontal line ($w_{eff}=-1/3$) shows the border of acceleration and deceleration. \label{fig:weff}}
\end{figure}

\begin{figure}[pb]
\centering
\includegraphics[width=9cm]{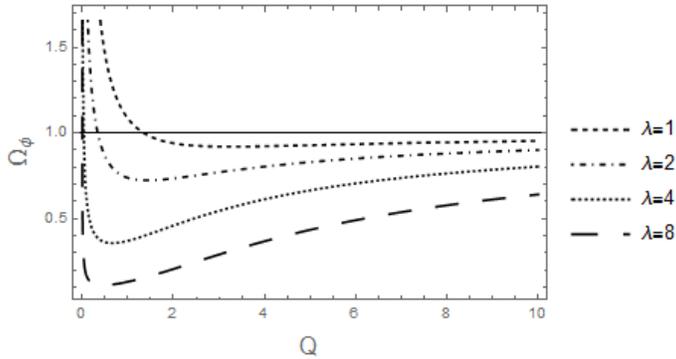}
\vspace*{8pt}
\caption{Density parameter of the dark energy $\Omega_\phi$ as a function of $Q$, for $\lambda=1$, $2$ and $4$. We show in the horizontal line the limit $\Omega_\phi=1$, which does not allow a range of values of $Q$. \label{fig:omegade}}
\end{figure}

In Figures \ref{fig:weff} and \ref{fig:omegade} we show the effective equation of state $w_{eff}$ and the density parameter for dark energy $\Omega_\phi$, respectively, as a function of $Q$, both for the point (g) (see Table \ref{criticalpoints}). Notice that depending on the value of $Q$, the universe does not exhibit an accelerated expansion for this fixed point, as we see from the horizontal line in Figure \ref{fig:weff}. The values of $Q$, at which they reach the upper limits of Eq. (\ref{restriction}) and $\Omega_\phi$, are, of course, the same. This feature  can also be checked from  Figures \ref{f1} and \ref{fig:omegade}. This fixed point is a tachyonic-dominated universe only for a range of values of $\lambda$ with specific $Q$. We have $\Omega_\phi=1$ for $\lambda=1$ and $Q\approx 1.35$, and for $\lambda=2$ and $Q\approx 0.35$, for instance. The effective equation of state for these two values  is $w_{eff}=-0.72$ and $-0.29$, respectively. Thus, the former $w_{eff}$ implies an accelerated expansion, while the latter $w_{eff}$ does not.

\subsection{Stability around the critical points} 

In order to study stability of the fixed points $(x,y, z)=(x_c,y_c,z_c)$, we consider linear perturbations $\delta x$, $\delta y$ and $\delta  z$ around them. The perturbations satisfy the following differential equations 

\begin{equation}\begin{aligned}\label{eq:perturbation}
\frac{d}{dN}\begin{pmatrix} \delta x \\ \delta y\\\delta z \end{pmatrix}= \begin{pmatrix} \frac{\partial f}{\partial x} & \frac{\partial f}{\partial y} &\frac{\partial f}{\partial \lambda z}\\ \frac{\partial g}{\partial x}& \frac{\partial g}{\partial y} & \frac{\partial g}{\partial z}\\ \frac{\partial h}{\partial x}& \frac{\partial h}{\partial y}& \frac{\partial h}{\partial z}\end{pmatrix}\begin{pmatrix} \delta x \\ \delta y\\\delta z\end{pmatrix}
\end{aligned}\end{equation}

\noindent where $f=f(x,y,z)$,  $g=g(x,y,z)$ and $h=h(x,y,z)$ are the right-hand side of (\ref{eq:dx/dn}), (\ref{eq:dy/dn}) and (\ref{eq:dz/dn}), respectively. The $3\times 3$ Jacobian matrix is evaluated at the fixed points and it posseses three eigenvalues $\mu_1$, $\mu_2$ and $\mu_3$. The general solution for the evolution of linear perturbations are 

\begin{equation}\begin{aligned}\label{soldeltax}
\delta x= Ae^{\mu_1N}+ Be^{\mu_2N}+Ce^{\mu_3N},\\
\delta y= De^{\mu_1N}+ Ee^{\mu_2N}+Fe^{\mu_3N},\\
\delta z= Ge^{\mu_1N}+ He^{\mu_2N}+Ie^{\mu_3N},
\end{aligned}\end{equation}

\noindent where the capital latin letters (A-I) are integration constants. The stability around the fixed points depends on the nature of the eigenvalues, in such a way that they are stable points if they have negative values ($\mu_1, \mu_2, \mu_3<0$), unstable points if they have positive values ($\mu_1, \mu_2, \mu_3>0$) and saddle points if at least one eigenvalue has positive (or negative) value, while the other ones have oposite sign. We show below the elements of the Jabobian matrix, using the notation $f_i\equiv \partial f/\partial i$

\begin{subequations}\label{jacobianelements}\begin{align}\label{eq:fx}
f_x=2x\left[3x-\sqrt{3}y\lambda+3Q\left(1-z^2-\frac{y^2}{\sqrt{1-x^2}}\right)\right]+(x^2-1)\left(3-\frac{3Qxy^2}{(1-x^2)^{3/2}}\right),\end{align}

\begin{align}\label{eq:fy}
f_y=(x^2-1)\left(-\sqrt{3}\lambda -\frac{6Qy}{\sqrt{1-x^2}}\right),\end{align}

\begin{align}\label{eq:fz}
f_z=-6Qz(x^2-1),\end{align}

\begin{align}\label{eq:gx}
g_x=\frac{y}{2}\left[-\sqrt{3}y\lambda+ 3x\frac{y^2}{\sqrt{1-x^2}}\right],
\end{align}

\begin{align}\label{eq:gy}
g_y=-\sqrt{3}xy\lambda- \frac{9}{2}y^2\sqrt{1-x^2}+\frac{1}{2}(3+z^2),
\end{align}

\begin{align}\label{eq:gz}
g_z=yz,
\end{align}

\begin{align}\label{eq:hx}
h_x=\frac{3xz}{2}\frac{y^2}{\sqrt{1-x^2}},
\end{align}

\begin{align}\label{eq:hy}
h_y=-3yz\sqrt{1-x^2},
\end{align}

\begin{align}\label{eq:hz}
h_z=z^2-2+\frac{1}{2}\left[\frac{3y^2}{\sqrt{1-x^2}}(x^2-1)+3+z^2\right].
\end{align}
\end{subequations}

Using Eq. (\ref{jacobianelements}) the eingenvalues of the Jacobian matrix were found for each fixed point in Table \ref{criticalpoints}.  The results are shown in Table \ref{stability}.

\begin{figure*}
\centering
	\includegraphics[scale=0.55]{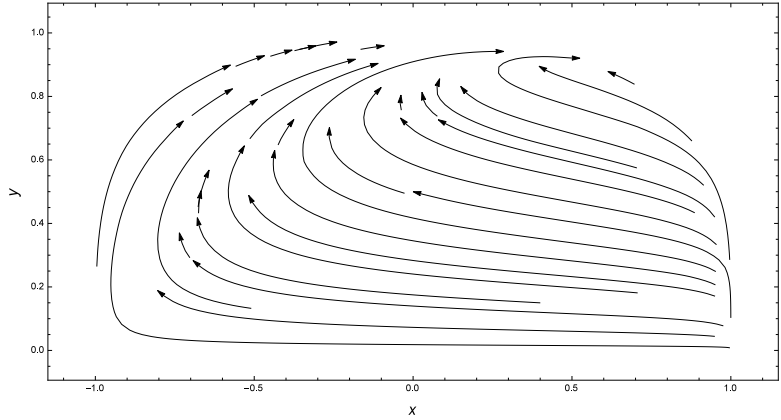}
	\caption{Phase plane for $\lambda=1$ and $Q=1.35$. The tachyonic-dominated solution (g) is a stable point at $x\approx 0.53$ and $y\approx 0.92$.}
		\label{fig:PhasePlane}
\end{figure*}

\begin{table}[ph]
\caption{Eigenvalues of the Jacobian matrix in Eq. (\ref{eq:perturbation}) and stability of the fixed points.}
{\begin{tabular}{@{}ccccc@{}} \toprule
  \text{Point} & $\mu_1$ & $\mu_2$ & $\mu_3$ &Stability\\
  \colrule
  (a)        & $6(1\pm Q)$& $\frac{3}{2}$& $-\frac{1}{2}$&   saddle\\
  (b)    & 6& 2& 1&   unstable\\
   (c)   & $-3 $& 4& 1&   saddle\\
  (d)   & $-3 $& $-3 $&$-2 $ &  stable\\
 (e)   & $\lambda^2y^2-3+\sqrt{3}Q\lambda y $&$\frac{\lambda^2y^2}{2}-3$ & $\frac{\lambda^2y^2}{2}-2$&  stable for $Q=0$ or $Q\lambda<0$\\ 
  (f)  & $3(Q^2-1)$& $\frac{3}{2}$& $-\frac{1}{2}$& saddle\\ 
  (g)   & $3\left(x^2-\frac{xy \lambda}{\sqrt{3}}\right)$& $\frac{3}{2}\left(\frac{xy \lambda}{\sqrt{3}}-2\right)$& $\frac{3}{2}\left(\frac{xy \lambda}{\sqrt{3}}-\frac{4}{3}\right)$& stable  \\
  \botrule
\end{tabular} \label{stability}}
  \end{table}

he first fixed point (a) is a saddle point, because the eigenvalues $\mu_2$ and $\mu_3$ have opposite signs, while $\mu_1$ can be either positive or negative. This point represents a matter-dominated universe, whose behavior is independent of the coupling $Q$. Both fixed points (b) and (c) lead to a  radiation-dominated universe, but the former is a unstable point while the latter is a saddle one. The point (d) is a late-time attractor and indicates an accelerated universe. The two last eigenvalues of the fixed point (e) are always negative. For this point $\lambda^2y^2=3x$, which in turn is between zero and three, then the first eigenvalue is also negative  if $Q=0$ or $Q\lambda<0$. Therefore, the point describes a dark-energy-dominated universe and can lead to a late-time accelerated universe  if the requirement  $\mu_1<0$ is statisfied. The effective equation of state depends only on $\lambda$, so the only thing the coupling $Q$ does, in the dynamical system analysis, is to change the property of the fixed point. The point (f) is  a saddle point with a matter-dominated behaviour. However, different from (a), the equation of state for the dark energy $w_\phi$ is no longer zero, but now depends on $Q$. Finally, the last fixed point (g) is stable, since $w_{eff}$ is between zero and minus one, as it is seen in Table \ref{criticalpoints}. Thus, $\mu_2$ and $\mu_3$ are always negative. The first eigenvalue is also negative because $\Omega_\phi\leq 1$. Therefore, the fixed can lead to a late-time accelerated universe, depending on the value of $\lambda$ and $Q$. Its behaviour is shown in Figure \ref{fig:PhasePlane}, for $\lambda=1$ and $Q=1.35$.

\section{Summary}\label{conclu}
The critical points showed in the last section describe the three phases of the universe, namely, the radiation-dominated era, the matter-dominated era, and the present dark-energy-dominated universe. The matter-dominated universe can be described by the two saddle points (a) and (f), whose difference among them is that the second one has a tachyonic equation of state that depends on $Q$. There are also two points that in principle can represent the radiation-dominated era: (b) and (c). The first one is unstable and the second one  is a saddle point  with equation of state for dark energy equals to minus one. A tachyonic-dominated universe is described by both points (d) and (e). Point (d) is stable and leads to an accelerated universe. Depending on the nature of the coulping and the potential, point (e) can also be stable, whose related equation of state depends on $\lambda$. Finally, the last fixed point (g) is stable and can describe an accelerated universe depending on the value of $\lambda$ and $Q$, and it is valid only for interacting dark energy. As we showed in the last section, there are options where the last point is viable to describe the present dark-energy-dominated era, although its equation of state is different from the  observed cosmological constant one \cite{planck2013cosmological}. 

Therefore, from the whole analysis, the transition radiation $\rightarrow$ matter $\rightarrow$ dark energy is cosmologically viable considering the following sequence of fixed points: (b) or (c) $\rightarrow$ (a) or (f) [$Q$ dependent] $\rightarrow$ (d), (e) [$\lambda$ dependent] or (g) [$Q$ and $\lambda$ dependent].

 Although the sequence is viable, the form of the potential dictates whether the fixed points are allowed or not. Among several possibilities in the literature, the potential $V(\phi)\propto \phi^{-n}$, for instance, leads to a dynamically changing $\lambda$ (either if $\lambda\rightarrow 0$ for $0<n<2$, or $\lambda\rightarrow \infty$ for  $n>2$) \cite{Abramo:2003cp}. A dynamically changing $\lambda$ is allowed for the fixed points (a)--(d) and (f), where for the point (d) $\lambda_c$ should be zero. On the other hand, points (e) and (g) require a constant $\lambda$, implying $V(\phi)\propto \phi^{-2}$. \cite{Gumjudpai:2005ry,micheletti2009,Padmanabhan:2002cp,Bagla:2002yn} Another possibility is a changing $\lambda$ which approaches asymptotically to a constant value \cite{landim2015ide}, thus, the case of constant $\lambda$ may be taken into account as an approximation. Overall, there can be a wide variety of potentials where the present dynamical analysis can be applied.

\begin{acknowledgments}
I thank Elcio Abdalla for comments. This work is supported by FAPESP Grant No. 2013/10242-1. 
\end{acknowledgments}

\bibliography{trab1}

\end{document}